\documentclass[fleqn,usenatbib,onecolumn]{mnras}

\usepackage[T1]{fontenc}

\DeclareRobustCommand{\VAN}[3]{#2}
\let\VANthebibliography\thebibliography
\def\thebibliography{\DeclareRobustCommand{\VAN}[3]{##3}\VANthebibliography}

\usepackage{graphicx}	
\usepackage{amsmath}	
\usepackage{epsfig} \usepackage{psfrag,epstopdf,color,float}
\usepackage{subfigure}

\title[A new model for spherical galaxies and its distribution functions]{A new model for spherical galaxies and its distribution functions}

\author[\text{{Z.} Jiang \& {Z.} Zhuo}]{
Zhenglu Jiang$^{1}$\thanks{E-mail: mcsjzl@mail.sysu.edu.cn} and 
Zanchen Zhuo$^{1}$\thanks{E-mail: zhuozch3@mail2.sysu.edu.cn}
\\
$^{1}$Department of Mathematics, Sun Yat-sen University, Guangzhou 510275, China\\
}

\date{Accepted 2025 February 24. Received 2025  February 20; in original form 2025  January 22}

\pubyear{2025}

\begin{document}
\label{firstpage}
\pagerange{\pageref{firstpage}--\pageref{lastpage}}
\maketitle
\graphicspath{{./Figure/}}

\begin{abstract}
A new potential is presented for spherical galaxies. 
The technique of the construction of our model is similar to that given by An and Evans. 
In a special case, its mass density becomes a special one of the Hernquist model.
Another special model is primarily discussed, and its intrinsic properties, such as velocity dispersions and surface densities, 
can be shown directly by numerical calculation. 
Its distribution functions in both isotropic and anisotropic cases can be expressed as some definite integrals which are 
suitable for numerical calculation.

\end{abstract}




\section{Introduction}
There is a long history to study the structure of the stellar system by determining its 
distribution functions (hereafter DFs) (\citealt{binney2008galactic}). 
It can be traced back to a very early time that Eddington (\citeyear{eddington1916distribution}) 
established the well-known Eddington formula for isotropic spherical galaxies:
\begin{equation}
	f(\mathcal{E})=\frac{1}{\sqrt{8}\pi^2}\frac{\mathrm{d}}{\mathrm{d}\mathcal{E}}
\int_{0}^{\mathcal{E}}\frac{\mathrm{d}\rho(\psi)}{\mathrm{d}\psi}\frac{\mathrm{d}\psi}{\sqrt{\mathcal{E}-\psi}},
\label{eq1}
\end{equation}
here $\mathcal{E}$ and $\psi$ represents the relative energy and the relative potential, respectively. 
For galaxy models with anisotropic velocity distributions, things become more complicated. 
However, there are also some profound results made by researchers 
({e.}{g.} \citealt{lynden1961stellar,hunter1975determination,osipkov1979spherical,dejonghe1986stellar,hunter1993two,jiang2007anisotropic}).

Usually, isotropic galaxy systems are first considered because it is simple to launch researches. 
Plummer (\citeyear{plummer1911problem}) established the well-known Plummer model which is of quite good properties in theoretical analysis. 
There are many impressive works ({e.}{g.} \citealt{nagai1976family,dejonghe1987completely}) about this model. 
Besides, a noteworthy achievement is de Vaucouleurs' $R^\frac{1}{4}$ model ({e.}{g.} \citealt{de1948recherches, binney1982phase}), 
which fits well the observations of many elliptical galaxies. 
However, this model is rarely used in theoretical analysis, 
because the deprojected mass density $\rho(r)$ cannot be solved analytically ({e.}{g.} \citealt{young1976tables}).
Based on Vaucouleurs' $R^\frac{1}{4}$ model, Jaffe (\citeyear{jaffe1983simple}) and 
Hernquist (\citeyear{hernquist1990analytical}) proposed a kind of model with different forms, 
which can be classified as $\gamma-$model (\citealt{dehnen1993family}). To take deeper consideration, 
Hernquist  (\citeyear{hernquist1990analytical}) gave a generalized density field defined by
\begin{equation}
	\rho(r)=\frac{C(\alpha,\beta,\gamma)}{4\pi a^3}\bigg(\frac{a}{r}\bigg)^\alpha\frac{1}{(1+(\frac{r}{a})^\beta)^\gamma}\label{eq2},
\end{equation} 
where $C(\alpha,\beta,\gamma)$ is a constant depending on $\alpha,\beta,\gamma$. 
It is found that  $\rho(r)\sim r^{-4}$ at large radii 
when the corresponding choices for $(\alpha,\beta,\gamma)$ are restricted by $\alpha+\beta\gamma=4$. 
In addition, the isochrone sphere ({e.}{g.} \citealt{henon1959amas,eggen1962evidence,evans1990flattened}) 
has attracted the attention of many astronomers, and its potential-density pair is as follows:
\begin{gather}
	\psi(r)=\frac{GM}{s+b},\label{isochronepsi}\\
	\rho(r)=\frac{M}{4\pi}\frac{b(2s+b)}{s^3(s+b)^2},\label{isochronerho}
\end{gather}
where, $s=\sqrt{r^2+b^2}$, $b$ is the scale length, $G$ and $M$ represent the gravitational constant and the total mass, respectively. 
This isochrone model was generalized by An and Evans (\citeyear{an2006galaxy}) replacing $s$ in (\ref{isochronepsi}) 
with  $s=(r^p+b^p)^\frac{1}{p}$ $ (p>0)$,  thus (\ref{isochronerho}) was changed as 
\begin{equation}
	\rho(r)=\frac{M}{4\pi}
\frac{2br^p+b^p(1+p)[(r^p+b^p)^\frac{1}{p}+b]}{r^{2-p}(r^p+b^p)^\frac{2p-1}{p}[(r^p+b^p)^\frac{1}{p}+b]^3}. \label{eq5}
\end{equation}

Inspired by the results mentioned above, we show a new galaxy model 
with the following dimensionless two-parameter potential:
\begin{equation}
	\psi(r)=\arctan\frac{1}{(r^p+b^p)^\frac{1}{p}}, \label{eq6}\\
\end{equation}
where 
$p$ and $b$ are two parameters of the model, $2\geq p>0$ and $b\geq 0$. 
According to Poisson's equation, we can get its density as follows:
\begin{gather}
	\rho(r)=\frac{1}{4\pi G}\frac{2r^p+b^p(1+p)[(r^p+b^p)^\frac{2}{p}+1]}
{r^{2-p}(r^p+b^p)^\frac{2p-1}{p}[(r^p+b^p)^\frac{2}{p}+1]^2},
\label{eq7}
\end{gather}
where $G$ is the gravitational constant. 

In (\ref{eq7}), if $p>2$, $\rho\rightarrow0$ as $r\rightarrow0$, which is unrealistic, so the $p$ to be considered is restricted in $(0,2]$. 
It is easy to see that the mass density (\ref{eq7}) with $b=0$ is the same as (\ref{eq2}) with $(\alpha,\beta,\gamma)=(1,2,2)$. 
 When $b=0$, $\rho(r)\sim r^{-1}$ near the centre, while  $\rho(r)\sim r^{-5}$ at large radii, 
just as that of the Plummer model. 
When $b>0$ and $0<p<2$, $\rho\sim r^{p-2}$ with a cusp at the centre,  while $\rho\sim r^{-p-3}$, 
 as $r\rightarrow+\infty$. When $b>0$ and $p=2$, $\rho$ is constant at the centre while $\rho\sim r^{-5}$ as $r\rightarrow+\infty$. 

The limiting model with $b=0$ is in the literature, and equals the $(1/2,5,1)$ model in Zhao's (\citeyear{zhao1996analytical}) paper 
or  the identical $(2,5,1)$ model of An and Zhao (\citeyear{an2013fitting}). 
It can be seen as a model that is intermediate between the Plummer model 
and the Wilkinson and Evans (\citeyear{wilkinson1999present}) model. 
A brief comparison with the properties of these three models is shown in {Figs.} \ref{fig1b}-\ref{fig3b}.

This paper is organized as follows. In Section 2, some isotropic DFs and surface densities are given in the case of $p=1$; 
In Section 3, the presented model is generalized to the anisotropic case, 
and the corresponding properties are given subsequently; Section 4 is a brief conclusion.

\section{Isotropic DF and Surface Density}
In this section, we show some isotropic DFs and surface densities of our model 
given by the potential-density pair (\ref{eq6}) and (\ref{eq7}) in the case of $p=1$. 

\subsection{Properties of isotropic DF}
For our convenience of numerical calculation, put $G=1$, here and everywhere below.
Corresponding to the mass density (\ref{eq6}), the cumulative mass distribution is given by 
\begin{equation}
	M(r)=-\frac{r^2}{G}\frac{\mathrm{d}\psi}{\mathrm{d}r}
=\frac{r^{p+1}(r^p+b^p)^{\frac{1}{p}-1}}{1+(r^p+b^p)^{\frac{2}{p}}} ,
\label{eq8}
\end{equation}
and the circular velocity is
\begin{equation*}
	v_c^2(r)=\frac{M(r)}{r}=\frac{r^p}{(r^p+b^p)^\frac{p-1}{p}[(r^p+b^p)^\frac{2}{p}+1]}.
\end{equation*}
With the help of (\ref{eq6}), $r$ can be expressed as 
\begin{equation}
	r=\bigg[\frac{1}{\tan^p (\psi)}-b^p\bigg]^\frac{1}{p}.
\end{equation}

\begin{figure}
	\centering
	\begin{subfigure}[]
	{
		\psfrag{E}{$\mathcal{E}$}
		\psfrag{log10f}{$\log_{10}f$}
		\includegraphics[scale=0.5]{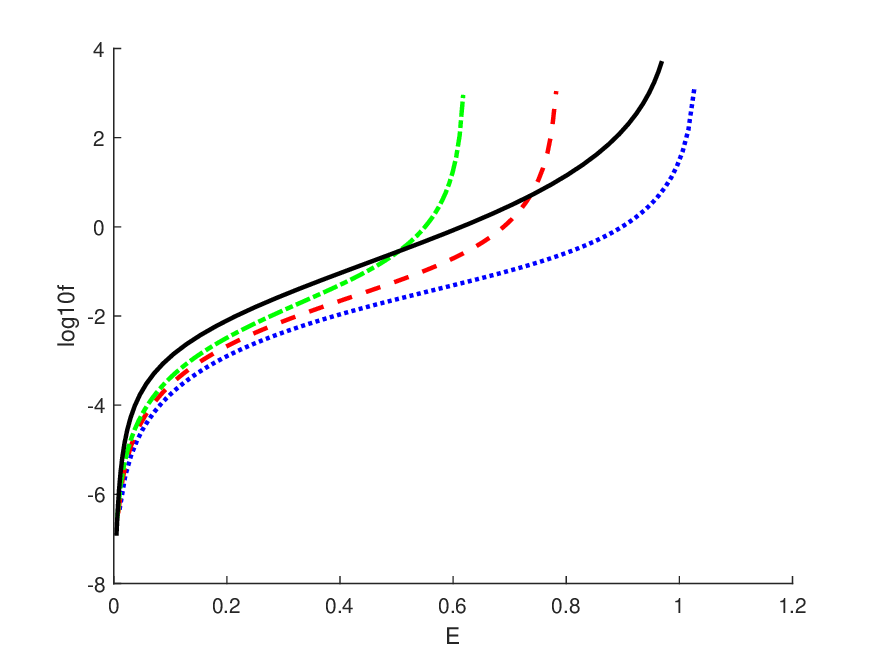}
		\label{fig1a}
	}
           \end{subfigure}
	\centering
	\begin{subfigure}[]
            {
		\psfrag{E}{$\mathcal{E}$}
		\psfrag{log10f}{$\log_{10}f$}
		\includegraphics[scale=0.5]{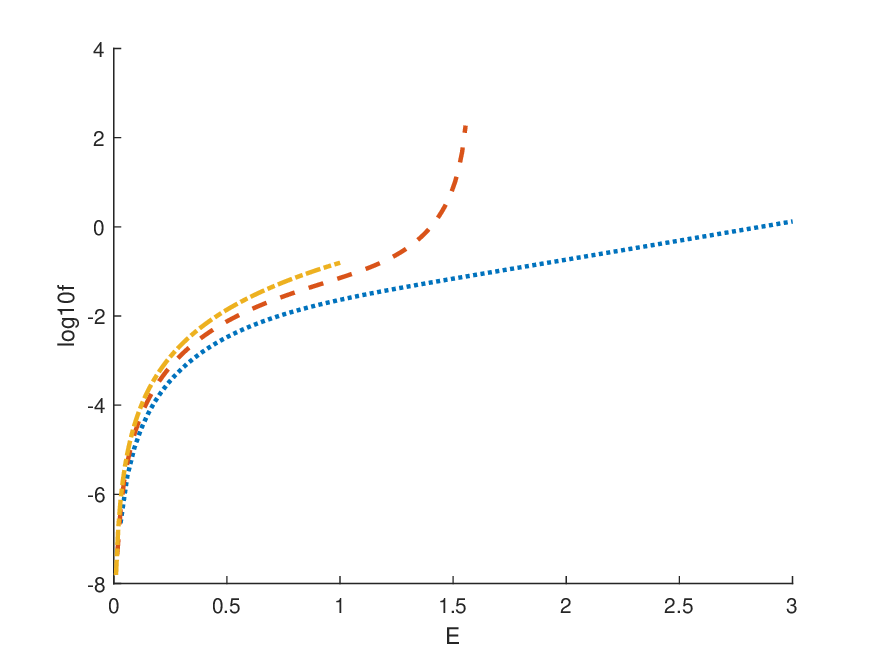}
		\label{fig1b}
	}
           \end{subfigure}
	\caption{(a) DFs for the $R^{\frac{1}{4}}$ model (solid curve), and for the presented model, 
		with different $b$ [$b=0.6$ (dotted), $b=1$ (dashed),  and $b=1.4$ (dashed-dotted)]. 
(b) DFs for our model with $b=0$ (dashed), the Wilkinson and Evans model (dotted), and the Plummer model (dashed-dotted).}
\label{fig1}
\end{figure}

In the following, we mainly consider the special case of $p=1$, that is, the potential-density pair given by
\begin{gather}
	\psi(r)=\arctan\frac{1}{r+b},\label{eq11}\\
	\rho(r)=\frac{1}{2\pi}\frac{br+1+b^2}{r\big[(r+b)^2+1\big]^2}. \label{eq12}
\end{gather}
By use of (\ref{eq11}),  (\ref{eq12}) can be expressed as a function of $\psi$:
\begin{equation}
	\rho(\psi)=\frac{1}{2 \pi}\frac{b+\tan (\psi)}{1-b\tan (\psi)}\sin^4(\psi). \label{eq13}
\end{equation}
Notice that  (\ref{eq13}) leads to the following property
\begin{equation*}
	\frac{\mathrm{d}\rho(\psi)}{\mathrm{d}\psi}\bigg|_{\psi=0}=0.
\end{equation*}
Thus, by Eddington's formula (\ref{eq1}),  
we can get the following DF
\begin{equation}
	f(\mathcal{E})=\frac{1}{8\sqrt{2}\pi^3}\int_{0}^{\mathcal{E}}
\frac{g(\psi)\tan^2(\psi)}{\cos (\psi)[1-b\tan (\psi)]^3}\frac{\mathrm{d}\psi}{\sqrt{\mathcal{E}-\psi}},\label{eq14}
\end{equation}
where
\begin{equation*}
	\begin{split}
	g(\psi)=&(9+b^2)\sin (\psi)+(7-b^2)\sin(3\psi)+2(1-3b^2)\sin(5\psi)\\
	&+b(7-b^2)\cos (\psi)+b(11+3b^2)\cos(3\psi)+2b(3-b^2)\cos(5\psi).
	\end{split}
\end{equation*}
{Fig.} \ref{fig1a} displays the different profiles of the isotropic DF (\ref{eq14}) with three choices of $b=0.6,1,1.4$, 
with a comparison to the $R^\frac{1}{4}$ model. When the energy $\mathcal{E}$ is near zero, 
the DFs of the presented model approximate to that of the $R^\frac{1}{4}$ model. 
The difference of DFs becomes larger as $\mathcal{E}$ is larger. 
In {Fig.} \ref{fig1b},  we use the unit dimensionless parameters for the  Plummer model and the Wilkinson and Evans model. 

\begin{figure}
	\centering
	\begin{subfigure}[]
	{
		\psfrag{log10r}{$\log_{10}r$}
		\psfrag{sigma2r}{$\sigma_r^2$}\includegraphics[scale=0.5]{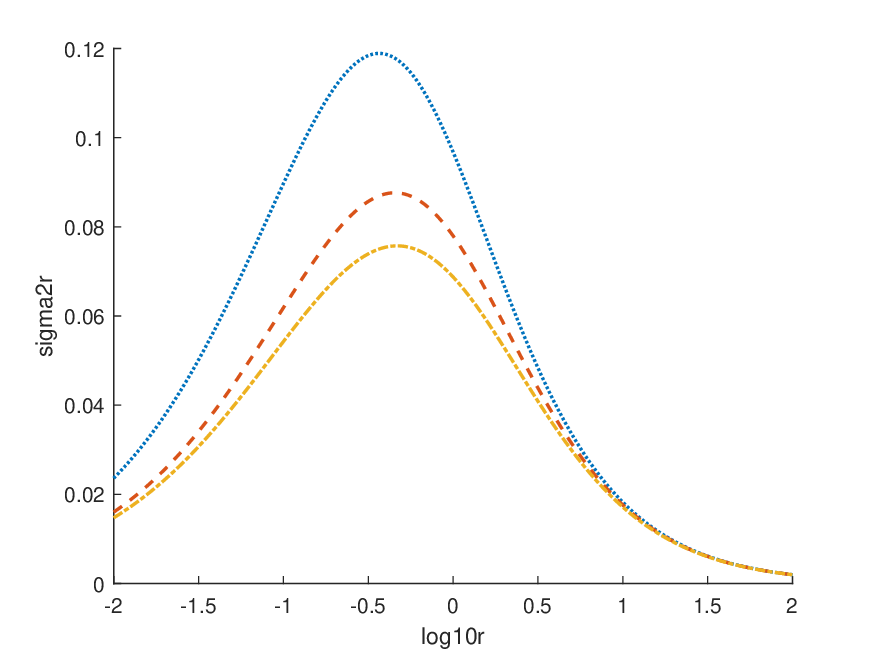}
		\label{fig2a}
	}
            \end{subfigure}
	\centering
	\begin{subfigure}[]
{
	\psfrag{log10r}{$\log_{10}r$}
	\psfrag{sigma2}{$\sigma_r^2$}
	\includegraphics[scale=0.5]{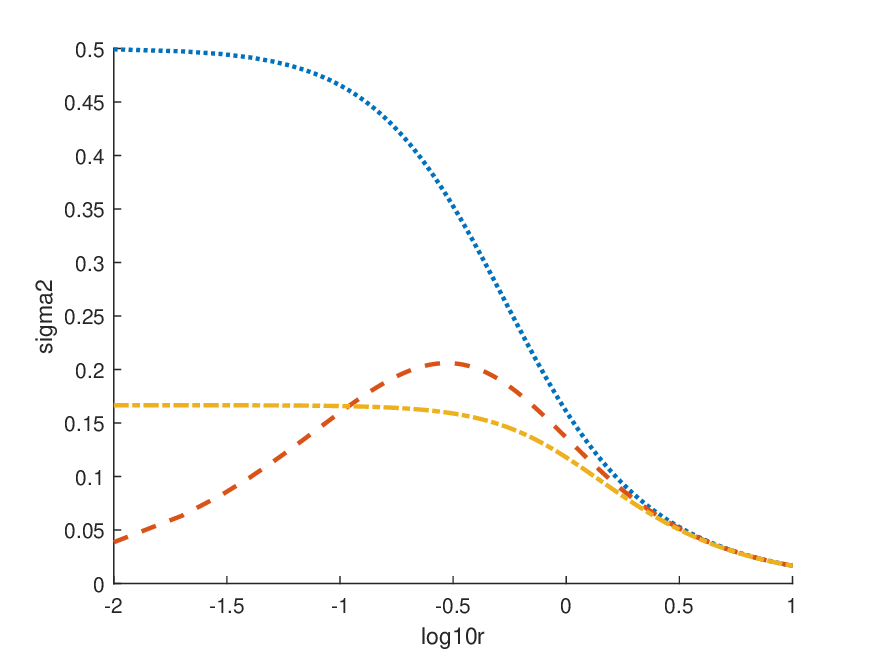}
	\label{fig2b}
}
            \end{subfigure}
\caption{(a) Velocity dispersions of the isotropic model given by (\ref{eq11}) and  (\ref{eq12}) 
	with different $b$ [$b=0.6$ (dotted), $b=1$ (dashed),  and $b=1.4$ (dashed-dotted)] as in {Fig.} \ref{fig1a}.
            (b) Velocity Dispersions for our model with $b=0$ (dashed), the Wilkinson and Evans model (dotted), 
            and the Plummer model (dashed-dotted).}
\label{fig2}
\end{figure}

For a non-rotating, spherical system,  the radial velocity dispersion $\sigma_r^2$ satisfies the Jeans equation
\begin{equation}
	\frac{1}{\rho}\frac{\mathrm{d}}{\mathrm{d}r}\big(\rho\sigma_r^2\big)+2\beta\frac{ \sigma_r^2}{r}=\frac{\mathrm{d}\psi}{\mathrm{d}r}, 
\end{equation}
where $\beta=1-\frac{1}{2\sigma_r^2}(\sigma_\theta^2+\sigma_\phi^2)$ is the anisotropy parameter of Binney,  and $\sigma_\theta^2$ 
is the angular velocity dispersion. It is well known that, for the isotropic system, 
$\sigma_r^2=\sigma_\theta^2=\sigma_\phi^2$ and so $\beta=0$, 
$\sigma_\phi^2$ is the angular velocity dispersion, too. 
The radial velocity dispersion $\sigma_r^2$ of our model given by (\ref{eq11}) and  (\ref{eq12}) can be expressed by 
\begin{equation}
\begin{split}
	\sigma_r^2\equiv \sigma_r^2(r)= 
\frac{r\big[(r+b)^2+1\big]^2}{2(1+b^2)^2(br+1+b^2)}
\bigg[\ln\frac{(r+b)^2+1}{r^2}-(3b+b^3)\arctan\frac{1}{r+b}\bigg] \\
-\frac{r}{4(br+1+b^2)}
-\frac{r\big[1-b(r+b)\big]\big[(r+b)^2+1\big]}{2(1+b^2)(br+1+b^2)}.
\label{eq15}
\end{split}
\end{equation}
{Fig.} \ref{fig2a} shows the profiles of the velocity dispersion (\ref{eq15}) with three choices of $b=0.6,1,1.4$. 
The velocity dispersions are finite everywhere, and vanish when $r\rightarrow0$ or $r\rightarrow+\infty$, since
\begin{equation}
	\sigma_r^2\sim r\ln\frac{1}{r}~(r\rightarrow0),~ \sigma_r^2\sim r^{-1}~(r\rightarrow\infty).
\end{equation}
The parameters in {Fig.} \ref{fig2b} are the same as in {Fig.} \ref{fig1b}. 
It is worth mentioning that, these profiles of the velocity dispersions in {Fig.} \ref{fig2a} 
resemble those of the spherical singular 
isothermal model (\citealt{osipkov1979simplest}).

\subsection{Surface Density}
\begin{figure}
	\centering
	\begin{subfigure}[]
	{
	\centering
	\psfrag{log10R}{$\log_{10}R$}
	\psfrag{log10SigmaR}{$\log_{10}\Sigma(R)$}
	\includegraphics[scale=0.5]{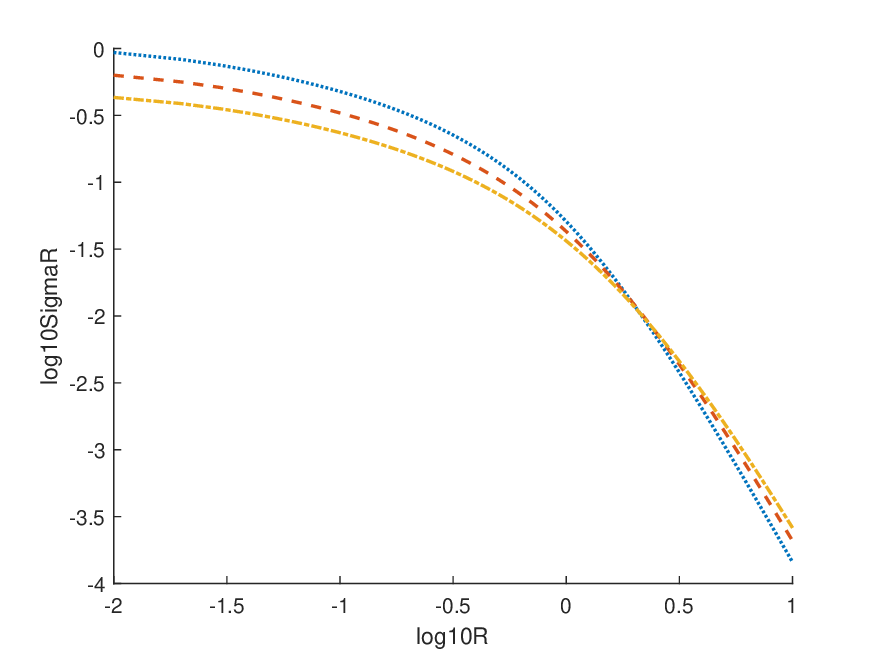}
	\label{fig3a}
	}
            \end{subfigure}
	\centering
	\begin{subfigure}[]
	{
	\centering
	\psfrag{log10R}{$\log_{10}R$}
	\psfrag{log10SigmaR}{$\log_{10}\Sigma(R)$}
	\includegraphics[scale=0.5]{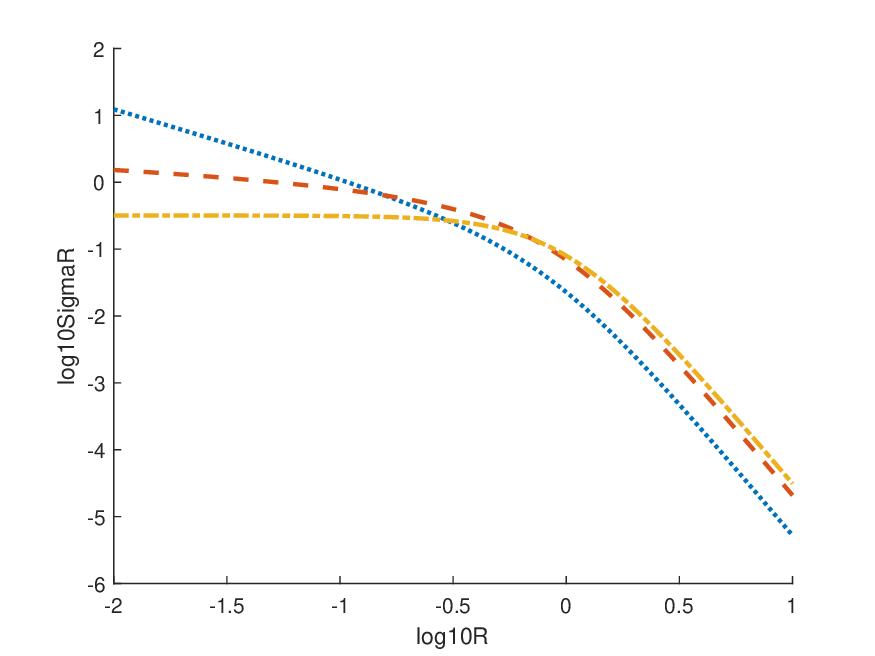}
	\label{fig3b}
	}
            \end{subfigure}
	\caption{(a) The profiles of the projected density (\ref{eq17}) with 
              the different parameters $b$ [$b=0.6$ (dotted), $b=1$ (dashed),  and $b=1.4$ (dashed-dotted)] as in {Fig.} \ref{fig1a}. 
(b) Surface Density for our model with $b=0$ (dashed), the Wilkinson and Evans model (dotted), and the Plummer model (dashed-dotted).}
\label{fig3}
\end{figure}

For the mass density (\ref{eq12}), the projected density is
\begin{equation}
\Sigma(R)=2\int_{R}^{+\infty}\frac{\rho(r)r}{\sqrt{r^2-R^2}}\mathrm{d}r
=\frac{8}{\pi }\int_{0}^{R}\frac{bt^3(t^2+R^2)+2(1+b^2)t^4}{\big[(t^2+2bt+R^2)^2+4t^2\big]^2}\mathrm{d}t.
\label{eq17}
\end{equation}
In (\ref{eq17}), the last equality is obtained by the Euler transform $\sqrt{r^2-R^2}=r-t$. 
When $R\rightarrow+\infty$, $\Sigma(R)\sim R^{-4}$.  
Although (\ref{eq17}) can be solved explicitly for given $b$, 
its general expression is absolutely complicated. 
But its numerical procedure is more desirable here. 
The profiles of (\ref{eq17}) are shown in {Fig.} \ref{fig3a}. 
The parameters in {Fig.} \ref{fig3b} are the same as in {Fig.} \ref{fig1b}. 

The cumulative surface density is
\begin{equation}
		S(R)=2\pi\int_{0}^{R}\Sigma(R')R'\mathrm{d}R'.
\end{equation}
Its numerical solution comes from (\ref{eq17}) naturally, 
and the effective radius $R_e$, which satisfies $S(R_e)=\frac{1}{2}$, 
can be attained numerically as well.

According to the book of Binney and Tremaine (\citeyear{binney2008galactic}), 
for a spherical, non-rotating system, the line-of-sight velocity dispersion $\sigma_l$
  is given by 
\begin{equation}
	\Sigma(R)\sigma_l^2(R)=2\int_{R}^{+\infty}(1-\beta\frac{R^2}{r^2})\frac{\rho\sigma_r^2 r}{\sqrt{r^2-R^2}}\mathrm{d}r.
\label{BTeq01}
\end{equation}
In the isotropic case, $\beta=0$ and  $\sigma_l^2$  is denoted by $\sigma_{l,i}^2$. Thus we can get 
\begin{equation}
	\sigma_{l,i}^2=\frac{2}{\Sigma(R)}\int_{R}^{+\infty}\frac{\rho\sigma_r^2r\mathrm{d}r}{\sqrt{r^2-R^2}},
\end{equation}
where $\rho$ and 
$\sigma_r^2(r)$ are given by (\ref{eq12}) and (\ref{eq15}), respectively.

 (\ref{BTeq01}) can be changed as 
\begin{equation}
	\Sigma(R)\sigma_l^2-R^2\int_{R}^{+\infty}\frac{\rho M(r)\mathrm{d}r}{r^2\sqrt{r^2-R^2}}
=\int_{R}^{+\infty}\bigg[2\rho\sigma_r^2+\frac{R^2}{r}\frac{\mathrm{d}(\rho\sigma_r^2)}{\mathrm{d}r}\bigg]\frac{r\mathrm{d}r}{\sqrt{r^2-R^2}}.
\label{BTeq02}
\end{equation}

In the case of purely circular orbits,  $\sigma_r^2=0$ and  $\sigma_l^2$  is denoted by $\sigma_{l,c}^2$. 
Thus, by (\ref{BTeq02}),  the line-of-sight velocity dispersion $\sigma_{l,c}^2$ is as follows:
\begin{equation}
	\sigma_{l,c}^2=R^2\int_{R}^{+\infty}\frac{\rho M(r)\mathrm{d}r}{r^2\sqrt{r^2-R^2}},
\end{equation}
where $\rho$ and $M(r)$ are given by (\ref{eq12}) and (\ref{eq8}) with $p=1$, respectively.

\begin{figure}
\centering
	\begin{subfigure}[]
	{
	\centering
	\psfrag{log10R}{$\log_{10}R$}
	\psfrag{sigmapiR}{$\sigma_{l,i}^2$}
	\psfrag{sigmapcR}{$\sigma_{l,c}^2$}
	\includegraphics[scale=0.5]{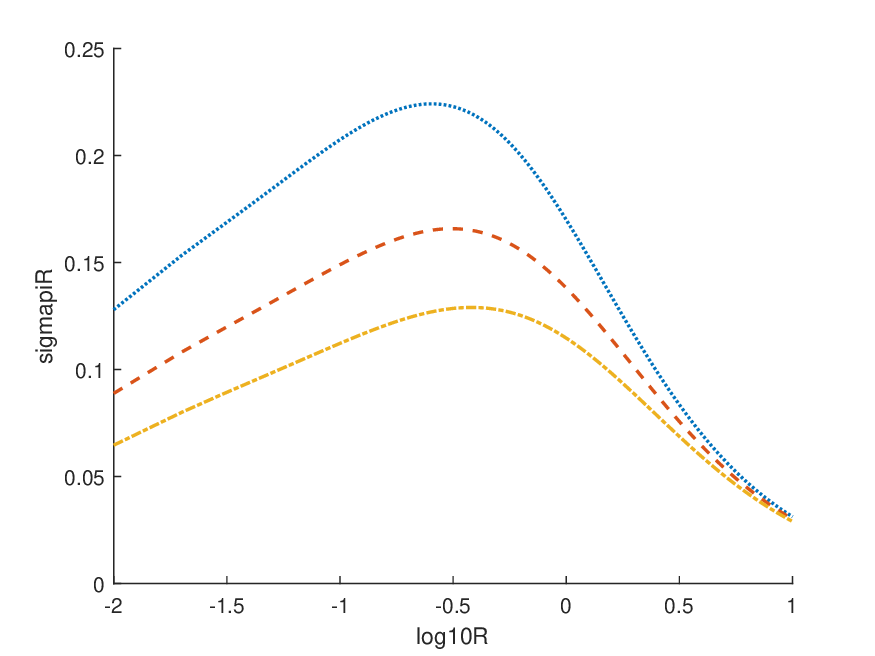}
	\label{fig4a}
	}
            \end{subfigure}
	\centering
	\begin{subfigure}[]
	{
	\centering
	\psfrag{log10R}{$\log_{10}R$}
	\psfrag{sigmapiR}{$\sigma_{l,i}^2$}
	\psfrag{sigmapcR}{$\sigma_{l,c}^2$}
	\includegraphics[scale=0.5]{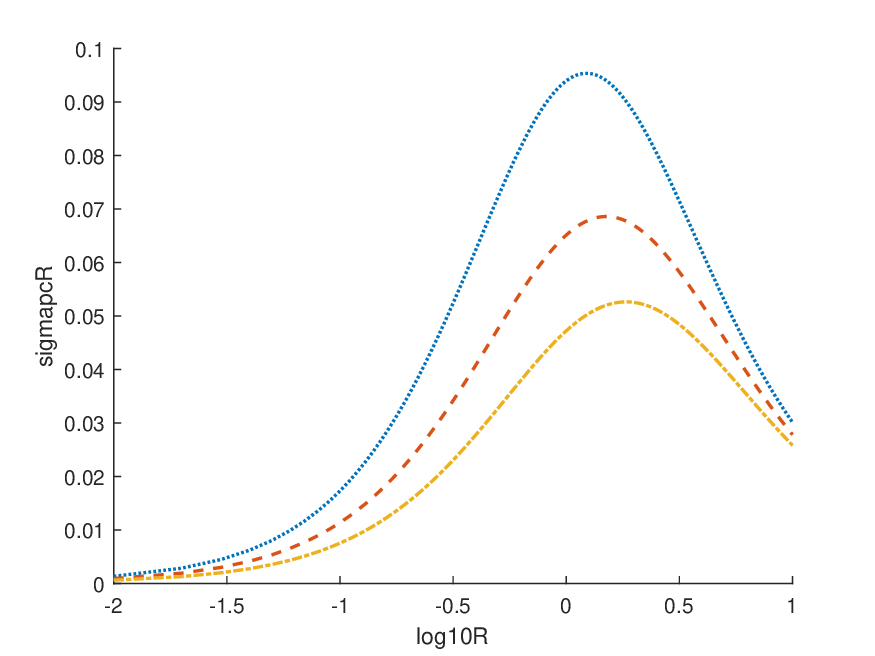}
	\label{fig4b}
	}
            \end{subfigure}
	\caption{The projected velocity dispersions in the two cases: (a) the isotropic case, and (b) the case of purely circular orbits. 
The format and the units of the curves are the same as in {Fig.} \ref{fig3a}. }
	\label{fig4}
\end{figure}

The profiles of $\sigma_{l,i}^2$ and $\sigma_{l,c}^2$ are shown in {Fig.} \ref{fig4}.

\section{Anisotropic Distribution Functions}
It is recognized that the anisotropic DF in the spherical system can be determined 
by two first integrals $\mathcal{E},L$,  that is, 
denoted by $f=f(\mathcal{E},L)$, 
where $\mathcal{E}$ and $L$ are the relative energy and 
the magnitude of the angular momentum $\bmath{L}$, respectively. 
It can be derived from the mass density of the form $\rho=\rho(\psi,r)$. 
Jiang and Ossipkov (\citeyear{jiang2007anisotropic}) provides some techniques to construct anisotropic DFs, and 
obtains the velocity dispersions $\sigma_r^2$ and $\sigma_T^2$.  A review about their solution is as follows. 
If the mass density $\rho=\rho(\psi,r)$ has a form of 
\begin{equation}
	\rho(\psi,r)=\sum_{n=0}^{m}\tilde{\rho}_n(\psi)r^{2n}, 
\label{jiangrho}
\end{equation} 
then a consistent DF corresponding to this density is
\begin{equation}
	f(\mathcal{E},L)=\sum_{n=0}^{m}h_n(\mathcal{E})L^{2n},
\label{jiangdf}
\end{equation}
where
\begin{equation}
	h_n(\mathcal{E})=\frac{1}{\sqrt{8}\pi^22^nn!}\bigg[\int_{0}^{\mathcal{E}}
\frac{\mathrm{d}^{n+2}\tilde{\rho}_n(\psi)}{\mathrm{d}\psi^{n+2}}
\frac{\mathrm{d}\psi}{\sqrt{\mathcal{E}-\psi}}+\frac{1}{\sqrt{\mathcal{E}}}
\bigg(\frac{\mathrm{d}^{n+1}\tilde{\rho}_n(\psi)}{\mathrm{d}\psi^{n+1}}\bigg)_{\psi=0}\bigg]. 
\label{eq22}
\end{equation}
For any fixed $\mathcal{E}$, $L_c$ is defined as the maximal value of angular momentum $L$, 
or it is denoted by $L_c(\mathcal{E})$. We can know that 
\begin{equation}
	L_c^2=\max\limits_{r\geq 0}\{2r^2\psi(r)-2r^2\mathcal{E}\}.
\label{eq23}
\end{equation}
That is, $f(\mathcal{E},L)\neq0$ only for those $(\mathcal{E},L)$ satisfying
\begin{equation*}
	\left\{
	\begin{array}{l}
		0\leq \mathcal{E}\leq \psi(0),\\
		0\leq  L\leq  L_c.
	\end{array}
	\right.
\end{equation*}

By (\ref{jiangrho}) and (\ref{jiangdf}),  the two velocity dispersions can be  given by 
\begin{gather}
	\sigma_r^2(\psi,r)=\frac{1}{\rho(\psi,r)}\sum_{n=0}^{m}r^{2n}\int_{0}^{\psi}\tilde{\rho}_n(\psi')\mathrm{d}\psi' ,\label{vdr}\\
	\sigma_T^2(\psi,r)=\frac{1}{\rho(\psi,r)}\sum_{n=0}^{m}(n+1)r^{2n}\int_{0}^{\psi}\tilde{\rho}_n(\psi')\mathrm{d}\psi'. \label{vdT}
\end{gather}

\begin{figure}
	\centering
	\psfrag{Eaxis}{$\mathcal{E}$}
	\psfrag{L2axis}{$L^2$}
	\includegraphics[scale=0.5]{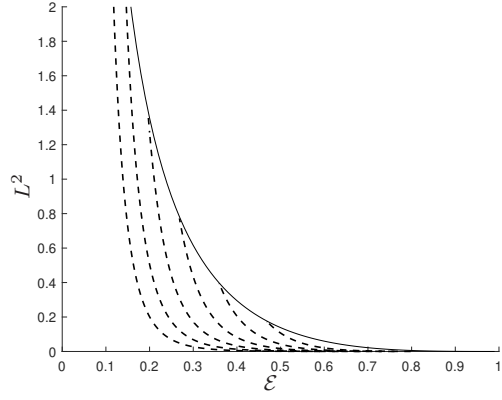}
	\caption{The contours of $f_1(\mathcal{E},L)$ given by (\ref{eq30}) with $b=\cot(1)$. 
The dashed curves are the contours and the solid curve is the boundary of the physical domain. 
Successive contour levels of $\log_{10}f_1(\mathcal{E},L)$ differ by factors of 0.4.}
	\label{fig5}
\end{figure}

To begin with, (\ref{eq12}) can be rewritten as the augmented mass density $\rho(\psi,r)$ in two simple forms given below. 
The first form of  (\ref{eq12}) is rewritten as
\begin{equation}
	\rho_k(\psi,r)=\frac{1}{2\pi}\frac{b+\tan (\psi)}{[1-b\tan (\psi)]^{2k+1}}\tan^{2k}(\psi)\sin^4(\psi) r^{2k}.
\label{eq20}	
\end{equation}
If $k$ is restricted to be a positive integer, those formulae given by Jiang and Ossipkov (\citeyear{jiang2007anisotropic}) are applicable. 
Using (\ref{jiangdf}), 
we can obtain the DF as follows:
\begin{equation}
	f_k(\mathcal{E},L)=\frac{L^{2k}}{\sqrt{8}\pi^22^kk!}\bigg[\int_{0}^{\mathcal{E}}
\frac{\mathrm{d}^{k+2}}{\mathrm{d}\psi^{k+2}}\tilde{\rho}_k(\psi)
\frac{\mathrm{d}\psi}{\sqrt{\mathcal{E}-\psi}}+\frac{\mathrm{d}^{k+1}\tilde{\rho}_k(\psi)}{\mathrm{d}\psi^{k+1}}
\bigg|_{\psi=0}\bigg],
\label{eq25}
\end{equation}
where
\begin{equation*}
	\tilde{\rho}_k(\psi)=\frac{1}{2\pi}\frac{b+\tan (\psi)}{[1-b\tan (\psi)]^{2k+1}}\tan^{2k}(\psi)\sin^4(\psi).
\end{equation*}
It could be a difficulty to obtain the derivative terms in (\ref{eq25}) when $k$ is large, but a simple case of $k=1$ can be provided as below 
(When $k=0$, the model is degenerated to be isotropic).  By (\ref{vdr}) and (\ref{vdT}), we can also get the following velocity dispersions:
\begin{gather}
	\sigma_{r,k}^2=\frac{[1-b\tan (\psi)]^{2k+1}}{[b+\tan (\psi)]\tan^{2k}(\psi)\sin^4(\psi)}
\int_{0}^{\psi}\frac{b+\tan (\psi')}{[1-b\tan (\psi')]^{2k+1}}\tan^{2k}(\psi')\sin^4(\psi')\mathrm{d}\psi', \\
	\sigma_{T,k}^2=(k+1)\sigma_{r,k}^2.
\end{gather}
Especially, when $k=1$, 
\begin{equation}
	f_1(\mathcal{E},L)=\frac{L^2}{4\sqrt{2}\pi^2}\bigg[\int_{0}^{\mathcal{E}}
P_1(\psi)  \frac{\tan^3(\psi)}{8\pi\cos^3(\psi)[1-b\tan (\psi)]^6} 
\frac{\mathrm{d}\psi}{\sqrt{\mathcal{E}-\psi}}\bigg], 
\label{eq30}
 \end{equation}
where
\begin{equation*}
	\begin{split}
	P_1(\psi)=&
(324+60b^2-8b^4)\sin (\psi)+(111+6b^2+15b^4)\sin(3\psi) \\
& +(31-54b^2-13b^4)\sin(5\psi)+(4-24b^2+4b^4)\sin(7\psi)\\
	&
+(228b-28b^3)\cos (\psi)+(156b+36b^3)\cos(3\psi)+(80b+8b^3)\cos(5\psi)+(16b-16b^3)\cos(7\psi).
	\end{split}
\end{equation*}
{Fig.} \ref{fig5} shows the contours of the logarithm of the DF given by (\ref{eq30}) with $b=\cot(1)$. 

\begin{figure}
	\centering
	\psfrag{E}{$\mathcal{E}$}
	\psfrag{L2}{$L^2$}
	\includegraphics[scale=0.5]{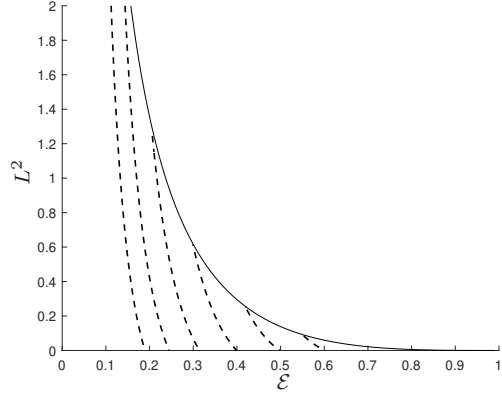}
	\caption{As in {Fig.} \ref{fig4}, but they are the contours of $f(\mathcal{E},L)$ 
given by (\ref{eq32}) with $b=\cot(1)$. Successive contour levels of $\log_{10}f(\mathcal{E},L)$ differ by factors of 0.3.}
	\label{fig6}
\end{figure}

The second form of (\ref{eq12}) can be changed as  
\begin{equation}
	\rho(\psi,r)=\frac{1}{2\pi}\frac{\sin^4(\psi)\tan (\psi)}{1-b\tan (\psi)}
\bigg[1+b^2+\frac{b\tan (\psi)}{1-b\tan (\psi)}r^2\bigg],
\label{eq31}
\end{equation}
and according to (\ref{jiangdf}), its corresponding DF is
\begin{equation}
	f(\mathcal{E},L)=h_0(\mathcal{E})+h_1(\mathcal{E})L^2,
\label{eq32}
\end{equation}
where
\begin{gather*}
	h_0(\mathcal{E})=\frac{1+b^2}{4\sqrt{2}\pi^3}
\int_{0}^{\mathcal{E}}
\frac{\big\{[9-7b\sin(2\psi)-4b\sin(4\psi)]+(9+2b^2)\cos(2\psi)+2(1-b^2)\cos(4\psi)\big\}\tan^3(\psi)}{[1-b\tan (\psi)]^3}\mathrm{d}\psi,\\
	\begin{split}
	h_1(\mathcal{E})=\frac{b}{8\sqrt{2}\pi^3}\int_{0}^{\mathcal{E}}
\frac{\tan^3(\psi)}{\cos^2(\psi)[1-b\tan (\psi)]^5}
\bigg[(4b^3-26b)\sin(2\psi)-(23b+5b^3)\sin(4\psi)+(2b^3-6b)\sin(6\psi)&\\
	+2(23+8b^2)\cos(2\psi)+2(7-2b^2)\cos(4\psi)+(2-6b^2)\cos(6\psi)+58-6b^2&
\bigg]\mathrm{d}\psi.
	\end{split}
\end{gather*}
{Fig.} \ref{fig6} shows the contours of the DF given by (\ref{eq32}) with $b=\cot(1)$. 

By (\ref{vdr}) and (\ref{vdT}), 
the velocity dispersions can be also given in Appendix \ref{appA}. 

\section{Conclusions}
We present a new family of spherical galaxy models with some interesting properties, 
and specifically display their DFs in a special case. 
The technique of the construction of our model is similar to that given by An and Evans (\citeyear{an2006galaxy}), 
and in the specific limiting case of $b=0$, 
its density becomes a special one of the Hernquist (\citeyear{hernquist1990analytical}) model, 
and equals the $(1/2,5,1)$ model in Zhao's (\citeyear{zhao1996analytical}) paper 
or  the identical $(2,5,1)$ model of An and Zhao (\citeyear{an2013fitting}). 
It can be easily seen that our arctangent potential leads to its mass density through Poisson's equation. Its mass density 
can be expressed as a function of the potential, thus Eddington's formula is suitable for constructing the isotropic DF $f(\mathcal{E})$. 
Unfortunately, $f(\mathcal{E})$ cannot be expressed in terms of elementary functions, 
but we can still make some research on its intrinsic properties with the help of numerical calculation. 
Our model approximates to de Vaucoulers' $R^\frac{1}{4}$ model closely 
when the energy is near zero. 
One of the characteristics of our model is that there are two free parameters, thus the model is widely applicable. 

The models with anisotropic properties are considered as well. 
The mass density is rewritten as two different forms, which are suitable for those formulae 
of Jiang and Ossipkov (\citeyear{jiang2007anisotropic}). 
Consequently the DFs and the velocity dispersions are established, and the contours of the DFs are displayed especially. 

According to the present observations, a constant-density core would be more 
prevalent (\citealt{tyson1998detailed}; \citealt{Wilkinson_2004}; \citealt{donato2004cores}), 
but the mass density with a typical cusp near the centre is considered to be very important as well 
(\citealt{navarro1995simulations}; \citealt{evans1997origin}; \citealt{moore1998resolving}). 
The model with $p=1$, considered in this paper, has  the latter property. When $p=2$, the model becomes the former one. 
It is rather complex, so we haven't made research on it yet. This model 
provides a test to study the structure and evolution of galaxies.  

\section*{Acknowledgements}
ZJ was supported by NSFC 10271121/11171356, and joint grants of NSFC 10511120278/10611120371 and RFBR 04-02-39026. 
The two authors would also like to thank the referee of this paper for his /her valuable comments in this work. 

\section*{Data availability}
The data underlying this article will be shared on reasonable request to the corresponding author.




\bibliographystyle{mnras}
\bibliography{reference} 




\appendix
\section{Velocity Dispersions of anisotropic DFs}
\label{appA}
We express the velocity dispersions of anisotropic DFs in terms of elementary functions.
According to  (\ref{vdr}) and (\ref{vdT}), the two velocity dispersions $\sigma_r$ and $\sigma_T$ of the mass density of the form (\ref{eq31}) 
can be expressed as follows:
\begin{gather}
	\begin{split}
		\sigma_r^2&
               =\frac{1}{2\pi\rho}\bigg[\frac{-(1+b^2)\mu(\psi)}{32(1+b^2)^3}+\frac{r^2\nu(\psi)}{64(1+b^2)^4(\cos(\psi)-b\sin(\psi))}\bigg],
	\end{split}\\
	\begin{split}
                       \sigma_T^2&
		=\frac{1}{2\pi\rho}\bigg[\frac{-(1+b^2)\mu(\psi)}{32(1+b^2)^3}+\frac{2r^2\nu(\psi)}{64(1+b^2)^4(\cos(\psi)-b\sin(\psi))}\bigg],
	\end{split}
\end{gather}
where  
\begin{gather}
	\begin{split}
	\mu(\psi)&=11+b(60\psi+14b+40b^2\psi+3b^3+12b^4\psi)
                              -4(3+4b^2+b^4)\cos(2\psi)+(1+b^2)^2\cos(4\psi)\\
		&+32\ln[\cos(\psi)-b \sin(\psi)]-8b(2+3b^2+b^4)\sin(2\psi)+b(1+b^2)^2\sin(4\psi),
           \end{split}\\
           \begin{split}
           \nu(\psi)&=-3b(1+b^2)^2(7+3b^2)\cos(3\psi)+b(1+b^2)^3\cos(5\psi)+4\cos(\psi)\bigg\{b (1+b^2)^2(5+2b^2)\\
		&+6(-5+15b^2+5b^4+b^6)\psi+96b\ln[\cos(\psi)-b\sin(\psi)]\bigg\}+3(1+b^2)^2(5+b^2)\sin(3\psi)-(1+b^2)^3\sin(5\psi)\\
		&-4\bigg\{-20+15b^2+7b^4(6+b^2)+6(-5+15b^2+5b^4+b^6)b\psi+96b^2\ln[\cos(\psi)-b\sin(\psi)]\bigg\}\sin(\psi).
	\end{split}
\end{gather}

\bsp	
\label{lastpage}
\end{document}